\documentclass[dvips]{acta}
\usepackage{supertabular,lscape,epsfig}
\usepackage{amssymb}
\usepackage{amsmath}
\usepackage{tablefootnote}
\usepackage{footnote}
\SetPages{0}{0}

\SetVol{0}{2019}

\begin{document}

\begin{Titlepage}
\Title{CCD Washington photometry of 10 open clusters or candidates projected close to the Galactic plane}

\Author{Clari\'a, J.\,J., Parisi, M.\,C., Palma, T., Ahumada, A.\,V.}{Universidad Nacional de C\'ordoba, Observatorio Astron\'omico, Laprida 854, C\'ordoba, Argentina and Consejo Nacional de Investigaciones Cient\'{\i}ficas y T\'ecnicas, CONICET, Godoy Cruz 2290, Ciudad Aut\'onoma de Buenos Aires, CPC 1425FQB, Argentina\\
e-mail:claria@oac.unc.edu.ar, celeste@oac.unc.edu.ar, tali@oac.unc.edu.ar, andrea@oac.unc.edu.ar}

\Author{Oviedo, C. G.}{Universidad Nacional de C\'ordoba, Facultad de Matem\'atica, Astronom\'ia y F\'isica, Argentina\\
e-mail:coviedo@oac.unc.edu.ar}

\Received{Month Day, Year}
\end{Titlepage}

\Abstract{We present high-quality CCD photometry in the Washington system C and T$_1$ passbands down to T$_1$  $\sim$ 19.5 mag in the fields of 10 Galactic open clusters (OCs) or candidates projected close to the Galactic plane, namely: ESO\,313-SC03, BH\,54, Ruprecht\,87, ESO\,129-SC32, BH\,217, Collinder\,347, Basel\,5, Ruprecht\,144, Archinal\,1 and Berkeley\,82. Four of these objects are located toward the Galactic centre direction within a solid angle of 21$^{\circ}$. No photoelectric or CCD photometry in the optical domain has been so far reported for five of these objects. Cluster radii are estimated from radial density profiles (RDPs) in the cluster fields. Using the cluster Washington (C-T$_1$,T$_1$) colour-magnitude diagrams (CMDs), statistically cleaned from field star contamination, we estimate reddening, heliocentric distance and age of the clusters by fitting Padova theoretical isochrones computed for the Washington system. In all cases, the best fittings were obtained with nearly solar metal content isochrones. Both RDPs and CMDs show that we are dealing with real OCs, except for Ruprecht 87 and Archinal\,1 that are found to be probably not physical systems. Differential reddening appears to be present across the fields of ESO\,313-SC03, ESO\,129-SC32, BH\,217, Collinder\,347 and Basel\,5. The studied OCs are located at d$_\odot$ = 1.0-5.0 kpc from the Sun and at Galactocentric distances $R_{GC}$ = 6.0-10.6 kpc, with mean reddening $E(B-V)$ in the range 0.10-1.30 mag and ages between 5 Myr (Collinder\,347) and $\sim$ 1000 Myr (Basel\,5). The estimated linear cluster radii are in the range of 0.4-3.2 pc. In general terms, the results obtained show fairly good agreement with previous photometric results. In some clusters, however, considerable differences are found between the present results and previous ones determined using near-infrared photometric data. The current study provides new OC parameters and some revisions to the OC catalogues.}
{{Galaxy: open clusters and associations: individual: ESO\,313-SC03, ESO\,129-SC32, BH\,54, Basel\,5, Ruprecht\,144, Archinal\,1, Ruprecht\,87, BH\,217, Collinder\,347, Berkeley\,82 - Techniques: photometric}
}

\section{Introduction}
Galactic open clusters (OCs) are fundamental tools to improve our understanding of the Galactic structure and of the chemical evolution of the Galactic disk (Friel 1995; Chen et al. 2003; Bonatto et al. 2006; Bobylev et al. 2007; Camargo et al. 2013). They are ideal objects for focusing on several astrophysical problems, such as star formation, stellar evolution and dynamic evolution of stellar systems (see, e.g., Bonatto and Bica 2005, and references therein). Particularly, OCs located toward the Galactic centre direction are relevant since they facilitate the tracing of the structure and evolution of the inner Galactic disk. Unfortunately, however, many OCs located toward the central parts of the Galaxy have not been studied in detail yet because they are affected by high interstellar absorption and/or by strong field star contamination. In this context, the near-infrared (NIR) photometric data gathered in the VVV survey (Minniti et al. 2010; Saito et al. 2012) happen to be very useful not only to discover new OCs (e.g., Borissova et al. 2011, 2014; Chen\'e et al. 2013; Ram\'{\i}rez Alegr\'{\i}a et al. 2014, 2016) and new globular cluster candidates (e.g., Minniti et al. 2017, and references therein) but also to determine astrophysical properties of extremely reddened objects projected onto the Galactic bulge (e.g., Palma et al. 2016). \\

This study is part of an observational project of Galactic OCs in the Washington photometric system, being currently developed at the Observatorio Astron\'omico de la Universidad Nacional de C\'ordoba (Argentina). Our goal is to determine reddenings, distances, ages, and if possible also metallicities for poorly known OCs (or candidates), preferably for those located toward the Galactic centre direction (see, e.g., Marcionni et al. 2014, hereafter M14). The Washington photometric system has been widely applied to young, intermediate-age and old clusters in the Galaxy (e.g., Geisler et al. 1992; Clari\'a et al. 2007; Piatti et al. 2009) and in the Magellanic Clouds (e.g., Geisler et al. 2003). Geisler et al. (1991) and Geisler and Sarajedini (1999) have demonstrated the pross offered by this system in the study of Galactic and/or extragalactic star clusters. \\

We have planned this paper in the following way. Section 2 describes the cluster sample, the observational setup and the reduction technique. The procedure carried out to estimate cluster radii from the radial density profiles is described in Section 3. We also present in this section the Washington (C-T$_1$,T$_1$) colour-magnitude diagrams (CMDs) and illustrate the determination of the studied cluster fundamental properties. The analysis and discussion of the results is presented in Section 4, while Section 5 summarizes our main conclusions.


\section{Cluster sample and data collection}
We selected from the New catalogue of Optically Visible Open Clusters and Candidates (Dias et al. 2002) ten relatively poorly studied Galactic OCs, giving preference to five objects (ESO\,313-SC03, ESO\,129-SC32, Basel\,5, Ruprecht\,144 and Archinal\,1) which, as far as we know, were never before observed photoelectrically or with CCD in the optical spectral range (Table 1). All the selected objects are low Galactic-latitude clusters ($\mid$b$\mid$ $<$ 5$^{\circ}$) that are often heavily contaminated and sparsely populated. Four of them are projected toward the Galactic centre within a solid angle of 21$^{\circ}$. Their main designations as well as the equatorial and Galactic coordinates taken from Archinal and Hynes (2003, hereafter AH03) are given in Table 1, together with the angular diameters D and Trumpler (1930) classes given by the same authors. Although some structural and photometric parameters were derived for these objects by means of Two-Micron All-Sky Survey (2MASS) photometry (Kharchenko et al. 2013), only five of them (BH\,54, Ruprecht\,87, BH\,217, Collinder\,347 and Berkeley\,82) were previously photometrically observed in the optical domain, while none of the selected clusters was previously observed in the Washington photometric system. Figs. 1-3 show schematic finding charts of the observed clusters' regions (13.6' $\times$ 13.6'). \\

The observational setup and the stellar photometry procedure are identical to those we used in M14. To sum up, the CCD images were obtained using the 0.9 m telescope at Cerro Tololo Inter-American Observatory (CTIO, Chile), equipped with a 2048 x 2048 pixel Tektronix CCD, during the nights of 2008 May 9 and 11. The detector used has a pixel size of 24 $\mu$m, producing a scale on the chip of 0.396'' pixel$^{-1}$ (focal ratio f/13.5) and a total field size of 13.6' x 13.6'. We used the Washington C (Canterna 1976) and Kron-Cousins R$_{KC}$ filters in order to be consistent with our previous studies (e.g., M14). Because the R filter has a much higher through-put than the standard Washington T$_1$ filter (Geisler 1996), it is posible to transform R magnitudes with high precision in order to lead to T$_1$ magnitudes. Standard stars of selected areas SA\,101, 107 and 110 from the list in Geisler (1996) were also nightly observed to standardize our photometry. The selected standard stars cover a wide range in colour. The weather conditions kept very stable at CTIO, with a typical seeing of 1.0''-1.2''.\\

Table 2 shows the log of the observations with filters, exposure times and airmasses. We refer the reader to M14 for a detailed description of the data acquisition and reduction procedure. The relationships between instrumental and standard magnitudes were obtained by fitting equations (1) and (2) of M14. Typical rms errors of these equations are 0.027 and 0.028. The resulting mean transformation coefficients a$_i$ and b$_i$ (i = 1, 2 and 3) together with their errors are listed in Table 3. We finally used about 15 standard stars.  \\

The whole data collected for each cluster are made up of a running star number, the CCD X and Y coordinates, equatorial coordinates, the derived T$_1$ magnitude and C-T$_1$ colour, and the photometric errors $\sigma$(T$_1$) and $\sigma$(C-T$_1$). We derived magnitudes and colours for a total of 1652, 369, 3090, 4125, 1462, 6444, 1743, 1453, 199 and 1127 stars in the fields of ESO\,313-SC03, BH\,54, Ruprecht\,87, ESO\,129-SC32, BH\,217, Basel\,5, Collinder\,347, Ruprecht\,144, Archinal\,1 and Berkeley\,82, respectively. For the reader to have a clear idea of the Washington data obtained for each cluster, we show their form and content in Table 4 for one of the studied clusters (ESO\,313-SC03). The whole clusters' database can be seen in the on-line version of the journal. \\

The T$_1$ magnitude and C-T$_1$ colour photometric errors as a function of T$_1$ for stars measured in the field of Berkeley\,82 is shown in Fig. 4. Such behavior is typical in our photometry. A quick inspection of Fig. 4 shows that most stars brighter than T$_1$ $\sim$ 18.5 mag have errors lower than 0.02 mag in the T$_1$  magnitude and lower than 0.10 mag in the (C-T$_1$) colour. Taking into account the variation of the photometric errors as a function of the T$_1$ magnitude, we rely on the precision of the morphology and position of the main cluster features in the (C-T$_1$,T$_1$) CMDs.

\section{Data analysis}

\subsection{Cluster radii}

As shown in Figs. 1-3, the OCs analysed in this study present diversified morphologies depending probably on their different evolutionary phases. To identify the most likely cluster members and minimize field star contamination in the CMDs, we estimated cluster radii from radial density profiles (RDPs) by performing star counts as in M14. The clusters' centres were determined with a typical standard deviation of $\pm$ 6''. They are marked by a cross in Figs. 1-3. The resulting RDPs are shown {\bf in Fig. 5}. Owing to the particular stellar distribution in the field of ESO\,129-SC32, it was not possible to perform either a reliable determination of its centre or to construct its RDP. The cluster radius ($r_{cl}$), generally used as an indicator of the cluster's size, is defined as the distance from the cluster centre where the RDP intersects the background level.
 In general terms, our conservative choice of the cluster radii maximizes the contrast between the cluster and its surrounding field stellar population. It is also helpful to measure the full width at half maximum (FWHM) of the RDP, which plays an important role in the construction of some clusters' CMDs. Table 5 lists the radii at the FWHM ($r_{FWHM}$) and the resulting cluster radii ($r_{cl}$) given in arcmin. The linear radii expressed in parsec were calculated using the corresponding heliocentric clusters' distances derived in Section 3.2.

\subsection{Cluster properties from CMDs}
CMDs are essential tools in determining the fundamental parameters of the OCs. However, field-star contamination is an important source of uncertainty, especially for low-latitude OCs and/or for those projected against the bulge. Our OCs are located in crowded disk regions close to the Galactic plane, and because of their low latitude, field stars contaminate the CMDs, especially at faint magnitudes and red colours. In particular, BH\,217, Collinder\,347, Basel\,5 and Ruprecht\,144 are located 
toward the Galactic centre within a solid angle of scarcely 21$^{\circ}$. {\bf Fig. 6} shows the (C-T$_1$,T$_1$) CMDs for all the observed stars in the 13.6' $\times$ 13.6' cluster fields. As shown in the different panels, most of the clusters' CMDs are strongly contaminated by field stars. Since the fundamental cluster parameters estimation requires to minimize the field-star contamination in the first place, we applied a statistical method described in detail in Piatti and Bica (2012).
Briefly, this procedure consists of selecting between one and four field regions at a distance of between two and four times the cluster radius for each cluster. Then, their respective (C-T$_1$,T$_1$) CMDs are obtained. The sizes of the areas of each field regions must be equal to the cluster area. Next, stars lying within different intervals of magnitude-colour [$\Delta$T$_1$,$\Delta$(C-T$_1$)] in the CMD of each selected region are counted.  This method includes variable intervals, depending on how populated the studied region is. Finally, the number of stars counted for each interval [$\Delta$T$_1$,$\Delta$(C-T$_1$)] in the CMD of the surrounding field region is subtracted from the number of stars of the cluster region. For more details see Piatti and Bica (2012). We show in the upper panels {\bf of Figs. 7-9} the (C-T$_1$,T$_1$) CMDs constructed using only stars with colour uncertainties smaller than 0.06 mag, which are located within the adopted cluster radii (Table 5). As seen in these panels, it is difficult in some cases to identify the cluster main sequences (MSs) with precision. The middle panels in {\bf Figs. 7-9} correspond to the CMDs of the equal area comparison fields, and the bottom panels show the resulting field star decontaminated (C-T$_1$,T$_1$) CMDs of the clusters wherein the main cluster features can be identified.\\

The most common procedure of fitting theoretical isochrones to the field decontaminated CMDs was applied to determine the $E(C-T_1)$ colour excess, the T$_1$-M$_{T_1}$ distance modulus, and the age and metallicity of the clusters. Regarding the theoretical isochrone sets, we chose the ones computed by the Padova group (Bressan et al. 2012) in steps of $\delta$log (age) = 0.05 dex. The metallicities of the studied clusters were not previously estimated. This is the reason why we began by adopting isochrones of solar metal content (Z = 0.019, [Fe/H] = 0.0) to carry out the fits and changed the isochrones' metallicity in steps of $\delta$[Fe/H] = 0.05 dex towards subsolar values in order to reach the best match. As a first step, we fitted the zero-age main sequence (ZAMS) to the cluster CMDs so as to derive colour excesses and distance moduli for each chosen metallicity. Next, using each of the derived [E(C-T$_1$),T$_1$-M$_{T_1}$]$_Z$ sets, we performed isochrone fits. As references features for the fittings, we considered the brightest magnitude in the MS, the bluest point of the turn-off and the locus of the red giant clump, provided it was visible. Finally, we selected the isochrone or range of isochrones that we considered matched the cluster features best. In all cases, we finally found the best fittings using solar metal content isochrones, estimating in $\pm$ 0.2 dex the error of the derived (solar) metallicity of the cluster. Table 6 lists the resulting cluster $E(B-V)$ colour excess, heliocentric distance and age together with their estimated uncertainties. The latter correspond to the shift allowed to isochrone fitting before a mismatch is clearly perceived by eye inspection. To relate both colour excesses and distance moduli, we used the relations $E(C-T_1)$ = 1.97 $E(B-V)$ and M$_{T_1}$ = T$_1$ + 0.58 $\times$ $E(B-V)$ - $(V-Mv)$ from Geisler (1996). To derive heliocentric distances we adopted R$_V$ = A$_V$/$E(B-V)$ = 3.0. The bottom panels of {\bf Figs. 7-9} show the best isochrone fits obtained for each individual cluster (solid lines), together with two additional isochrones bracketing the derived age (dashed lines). As an example, we show in {\bf Fig. 10} two fittings of the CMDs of Berkeley 82 with isochrones of Z = 0.019 ([Fe/H] = 0.0) and Z = 0.013 ([Fe/H] = -0.2), respectively. Differences in these fittings are scarcely perceptible.\\

We also list in Table 6 the Galactocentric rectangular coordinates X$_\odot$, Y$_\odot$ and Z$_\odot$ and the Galactocentric cluster distance $R_{GC}$ derived by assuming 8.5 kpc for the Galactocentric distance of the Sun. As in Lyng\aa\,(1982), we adopted a heliocentric reference system with the X and Y axes lying on the Galactic plane and the Z-axis perpendicular to this plane. X points in the direction of the Galactic rotation, being positive in the first and second Galactic quadrants, while Y points toward the Galactic anticentre, being positive in the second and third quadrants. Finally, Z is positive toward the north Galactic pole. Previous $E(B-V)$ colour excesses, heliocentric distances and ages determined by different authors for the present cluster sample are listed in Table 7, for comparison purposes.

\section{Results}

We reported in Oviedo et al. (2017) some preliminary findings derived from Washington photometry for only three of the clusters here analized (Basel\,5, Collinder\,347 and Berkeley\,82). Those results may be considered only a first approach to the estimation of their fundamental parameters. Using NIR data taken from the 2MASS catalogue (Skrutskie et al. 2006), Bukowiecki et al. (2011, hereafter B11) derived basic parameters for ESO\,313-SC03, BH\,54, BH\,217, Collinder\,347, Basel\,5 and Archinal\,1. Through a combination of uniform kinematic and NIR photometric data included in the all-sky catalogue PPMXL (Roeser et al. 2010) and the 2MASS catalogue, Kharchenko et al. (2013, hereafter K13) reported some structural properties and fundamental parameters for more than 2800 mostly genuine OCs, among which are all the OCs of the present study. Nevertheless, due to the limitations posed by distance and reddening, the determination of the faintest clusters' parameters depends just on a few stars. Besides, note that using 2MASS data is reliable for well defined and clearly distinguishable OCs, which is not always the case. Indeed, once having made a careful analysis of the major large scale OC homogeneous parameter databases, Netopil et al. (2015) detected some trends and/or relevant offsets. Following these authors, the databases can sometimes present more than 20\% of problematic objects, which coud reduce their usefulness.\\

A brief description of the studied OCs is given in the following subsections, as well as a comparison of our results obtained from Washington photometry with those results of previous studies.\\

\subsection{ESO\,313-SC03}
This small group of faint stars in Vela, also designated as MWSC 1499 (Kharchenko et al. 2012, hereafter K12), was first recognized as a possible OC by Lauberts (1982). AH03 refer to it as belonging to Trumpler class II3, i.e., a detached cluster with a medium range in the brightness distribution of the stars (Fig. 1). As far as we know, no previous optical photoelectric o CCD photometric study exists for this object. The presence of the cluster is clearly seen in the RDP (Fig. 5), from which we estimate an angular radius of 2.1'. The cluster and field CMDs are not very different from each other {\bf (Fig. 7)}, the only clear difference being probably the distinct turn-off at the bright end. The cluster MS extends along $\sim$ 2.0 mag down to the turn-off at T$_1$ $\sim$ 16.5. There are a few subgiants and/or red giants in the cluster CMD, some of which may not be members. Although the broadness of a cluster MS is certainly due to several effects (binarity, rotation, evolution, etc.), in the case of ESO\,313-SC03, there seems to be a variation of the interstellar reddening across its field. The lower limit estimated by Burki (1975) for clusters with differential reddening is $\Delta(B-V)$ = 0.11 mag, equivalent to $\Delta(C-T_1)$ = 0.22 mag, if the ratio $E(C-T_1)/E(B-V)$ = 1.97 given by Geisler (1996) is adopted. From the corresponding bottom panel of {\bf Fig. 7}, we estimate $\Delta(C-T_1)$ = 0.6-0.7 mag, a value which largely exceeds the Burki's (1975) limit. A nearly similar amount of differential reddening seems to be present in the fields of ESO\,129-SC32 (Section 4.4), BH\,217 (Section 4.5), Collinder\, 347 (Section 4.6) and Basel\,5 (Section 4.7), while in case this effect existed in the remaining clusters it is by far less evident. Our most likely isochrone fit to the cleaned cluster CMD {\bf (Fig. 7)} suggests that ESO\,313-SC03 is a moderately young cluster (0.4 $\pm$ 0.1 Gyr), located about 5.0 kpc from the Sun, and affected by a large reddening $E(C-T_1)$ = 1.58 $\pm$ 0.20 mag, equivalent to $E(B-V)$ = 0.80 $\pm$ 0.10 mag (Table 6). Surprisingly, B11 derived a much smaller colour excess $E(B-V)$ = 0.19 mag , a smaller heliocentric distance $d$ = 3.71 kpc and a much older age of 2.8 Gyr (Table 7). More recently, however, K13 reported $E(B-V)$ = 1.08 mag, $d$ = 4.27 kpc and an age of 398 Myr. Although the results by B11 and K13 were derived using NIR data taken from the 2MASS catalogue, the discrepancies are evident. Our results show better agreement with those of K13, although our photometric study in the optical domain suggests a smaller cluster reddening. In particular, the large age difference between ours and B11's may be due to a difference in the turn-off point adopted in the CMD. \\

\subsection{BH\,54}
First recognized as an OC by van den Bergh and Hagen (1975), this small-sized object (IAU designation C0848-422), also known as MWSC 1575 (K12), seems to be a detached, moderately poor OC wherein most of the stars exhibit nearly the same apparent brightness (Fig. 1). van den Bergh and Hagen (1975) described it as a faint, very poorly populated object having a radius of 1.5', clearly visible both on red and on blue plates. The RDP suggests a dense core and a small $r_{cl}$ value of 2.1' (Table 5) in good agreement with AH03. As shown in {\bf Fig. 7}, the CMD of BH\,54 exibits a relatively narrow and well-defined MS extending along $\sim$ 5.5 mag without a clear evidence of some evolution. The bright MS in the cluster CMD is not seen in the field CMD so that most of the stars in this sequence are very probably cluster members. The best isochrone fit to the decontaminated cluster CMD {\bf (Fig. 7)} indicates that BH 54 is a relatively nearby ($\sim$ 1 kpc) very young cluster (32 Myr), affected by E(B-V) = 0.75 $\pm$ 0.03 mag, located very close to the Galactic plane (Table 6). These results show fairly good agreement with those previously derived by B11 and K13 using NIR photometric data as well as with those determined by Piatti et al. (2010) based on UBVI photometry (Table 7).

\subsection{Ruprecht\,87}
This is a small cluster candidate (C1013-504) located in a fairly rich Galactic field in Vela (Fig. 1). Also referred to as MWSC\,1783 by K12, Ruprecht\,87 is located at a somewhat higher Galactic latitude than the other clusters (Table 1). It shows the typical morphology of a Trumpler class III-2p, i.e., a poor, detached cluster candidate with little central concentration of member stars and a medium range of bright stars. Although rather poor, this cluster candidate stands out from the somewhat sparse field (Fig. 1). The presence of this cluster is clearly seen in the RDP (Fig. 5), from which we derive a radius of 1.7', in good agreement with the value of 2.0' reported by both AH03 and Lyng\aa\,(1987). However, since by inspecting {\bf Fig. 7} we do not see any clear sequence in the cleaned CMD, we conclude that Ruprecht\,87 is probably not a physical system but a random concentration of field stars. As for ESO\,313-SC03, this cluster candidate has not been previously observed photoelectrically or with CCD in the optical spectral region.\\

\subsection{ESO\,129-SC32}
Like most of the clusters recognized by Lauberts (1982), ESO\,129-SC32, also named MWSC\,1970 (K12), has not yet been the object of previous photometric studies in the optical spectral range. This cluster is located in a rich Galactic field, visible as a marginal enhancement in the stellar density from the field (Fig. 1). As stated in Section 3.1, due to the particular stellar distribution in the cluster field, it was not possible to perform either a reliable determination of its centre or construct its RDP. We then adopted the value of 2.0' estimated by AH03 to clean the cluster CMD. The cluster and field CMDs are not very different, so it is difficult to clearly identify the cluster MS {\bf (Fig. 8)}. There appears to be a clear variation of the interstellar reddenng across the cluster field, as in  ESO\,313-SC03. From the bottom-left panel of {\bf Fig. 8}, we estimate in $\Delta(C-T_1)$ = 0.7 mag the broadness of the cluster MS, mostly due to differential reddening. The cleaned CMD is relatively well matched by isochrones indicating a moderate young [log(age) = 8.5-8.7] cluster, located at an approximate distance of 2.00 $\pm$ 0.22 kpc with low reddening [$E(B-V)$ = 0.10 $\pm$ 0.09 mag] (Table 6). However, K13 estimated a much higher reddening of $E(B-V)$ = 0.42 mag, an identical cluster age and a somewhat smaller heliocentric distance of 1.56 kpc. \\

\subsection{BH\,217}
This is a very detached, moderately poor and relatively faint group of stars (Trumpler class I-2m), first recognized as an OC in Scorpius by van den Bergh and Hagen (1975). From the RDP (Fig. 5) we estimate a radius of 3.5' (Table 5), somewhat larger than the AH03 value, although slightly lower than that of 4.0' reported by Lyng\aa\,(1987). BH\,217 (C1712-407) or MWSC\,2553 (K12) is also projected onto a rich Galactic field (Fig. 2). The cluster CMD shows a rather broad MS probably due to severe foreground reddening {\bf (Fig. 8)}. As in BH\,54, the bright MS stars in the cluster CMD are not seen in the field CMD. Therefore, these are most probably cluster members. From the corresponding bottom panel of {\bf Fig. 8}, we estimate $\Delta(C-T_1)$ = 0.8-0.9 mag, clearly larger than the Burki's (1975) limit. As for ESO\,313-SC03 and ESO\,129-SC32, the existence of differential reddening in the field of BH\,217 makes the determination of the cluster parameters somewhat more uncertain than in the other clusters, particularly its heliocentric distance (Table 6).  Nevertheless, our results show in general terms a reasonable agreement with previous ones derived from photometric and spectroscopic data (Table 7).\\

\subsection{Collinder\,347}
This object (C1743-292), also known as BH\,244 (van den Bergh and Hagen 1975) or MWSC\,2688 (K12), is a relatively faint cluster first recognized in Sagittarius by Collinder (1931). AH03 refer to this object as belonging to Trumpler class II-2m,n i.e., a moderately rich, detached cluster associated with nebulosity, having little central concentration and medium-range bright stars (Fig. 2). It is projected onto the central bulge of the Galaxy almost exactly toward the Galactic centre direction (Table 1). Both Lyng\aa\,(1987) and AH03 report a comparatively large angular radius of 5.0', slightly smaller than the $r_{cl}$ value derived in this study (Table 5). \\

Given that Collinder\,347 has a comparatively larger radius (Table 5), it is not possible to apply to it the field-star decontamination process described in Section 3.2. In order to build an optimum CMD for this cluster, which represents a compromise between maximizing the number of cluster stars and minimizing the field star contamination, we performed different circular extractions around the cluster centre as shown in {\bf Fig. 11}.  Only stars for which color uncertainties are smaller than 0.06 mag are included in these diagrams. We began with the CMD for stars distributed within $r$ $<$ $r_{FWHM}$ as the cluster fiducial sequence reference (upper-left panel). Then, we varied the distance from the cluster centre starting at $r_{FWHM}$ and built different series of extracted CMDs. Finally, we chose the cluster CMD for stars distributed within $r$ $<$ $r_{clean}$ (upper-right panel), where $r_{clean}$ = 3.0' is the estimated radius which yields the best enhanced cluster fiducial features. Since we believe that this CMD maximizes the cluster population and minimizes the field star contamination, we used it to apply the isochrone fitting procedure. As shown in the two upper panels of {\bf Fig. 11}, the most prominent feature is a broad MS extending roughly from T$_1$ = 12.0 mag down to 18.0 mag. The width of this MS is clearly not the result of photometric errors, since these hardly reach a tenth of magnitude at any T$_1$ level (Fig. 4). As for BH\,217, Collinder\,347 presents a clear variation of the interstellar reddening across its field. There is a tendency for the $E(B-V)$ colour excesses to increase from east to west (Moffat and Vogt 1975). From the upper-right panel of {\bf Fig. 11}, we estimate in $\Delta(C-T_1)$ $\sim$ 0.6-0.7 mag the broadness of the cluster MS, mostly due to differential reddening. The same as for ESO\,313-SC03, ESO\,129-SC32 and BH\,217, the resulting parameters for Collinder\,347 present associated errors larger than those for the other clusters (Table 6). The lower-left panel of {\bf Fig. 11} shows the CMD for stars distributed within $r$ $<$ $r_{cl}$, while the lower-right panel is the CMD we adopted for the field. In conclusion, Collinder\,347 is a very young cluster in the Vela constellation affected by a significant amount of differential reddening. Previous photometric determinations of the mean cluster reddening, distance and age are fairly consistent with the present results (Table 7).\\

\subsection{Basel\,5}               
This cluster was first reported by Svolopoulos (1966) who observed it photographically in the RGU system deriving a heliocentric distance of 850 pc. As far as we know, these are the only photometric observations in the optical range. Basel\,5 is located very close to Collinder\,347, so it is also in a very rich field projected onto the central bulge of the Galaxy (Fig. 2). Also designated MWSC\,2727 (K12), Basel\,5 (C1749-300) is a very loose concentration of comparatively faint stars in Scorpius. AH03 refer to this object as belonging to Trumpler class III-2m,n, i.e., a moderately rich cluster associated with nebulosity, having little central concentration and medium-range bright stars (Fig. 2). The RDP around this cluster clearly shows an excess of stars in the cluster region {\bf (Fig. 5)}. According to both Lyng\aa\,(1987) and AH03, Basel\,5 has a small angular radius of 2.5', in very good agreement with our $r_{cl}$ value (Table 5). As shown in {\bf Fig. 8}, the cluster and the field CMDs are not very different from each other. Nevertheless, the cluster cleaned CMD exibits a well-defined MS extending along $\sim$ 3 mag, with evidence of some evolution. There are a few subgiants and/or red giants in the cluster CMD, some of which may be members. Most of the bright MS stars in the field with T$_1$ $<$ 15.0 mag and (C-T$_1$) $<$ 1.6 mag are not seen in the cleaned cluster CMD, so that they are not likely to be cluster members. Differential reddening of $\Delta(C-T_1)$ = 0.5 mag is estimated in the cluster field. The best isochrone fit to the decontaminated cluster CMD {\bf (Fig. 8)} indicates that Basel\,5, located about 2.51 kpc from the Sun, is a $\sim$ 1 Gyr cluster affected by $E(B-V)$ = 0.48 $\pm$ 0.15 mag (Table 6). The association of Basel\,5 with nebulosity is not expected to be found at the age of this cluster, so its Trumpler type may be due to the presence of a foreground nebulosity not associated with the cluster. Our results are somewhat different from those derived by K13 and significantly in discrepancy with those reported by B11 (Table 7).

\subsection{Ruprecht\,144}
Ruprecht\,144 (C1830-114) or MWSC\,2950 (K12) is one of the most poorly defined objects of the present sample. AH03 classified it as belonging to class IV-1p, i.e., a poor cluster in Scutum without any noticeable concentration having stars in a narrow range of brightness (Fig. 2). The new coordinates for the cluster centre derived in the current study are $\alpha$ = 18$^h$ 33$^m$ 34$^s$, $\delta$ = -11$^{\circ}$ 24' 57'', i.e., they differ by $\sim$ 15'' in right ascension and by $\sim$ 14'' in declination from those listed in the Dias et al. (2002) catalogue. The cluster RDP obtained using the new cluster centre shows that Ruprecht 144 has a core and a small radius $r_{cl}$ = 1.0' (Table 5), in excellent agreement with the value reported by AH03. As far as we know, no previous optical photometric data exist for this sparse group of stars. {\bf Fig. 9} shows that the cluster CMD is very different from the field CMD. A narrow, well-defined cluster MS extending along $\sim$ 6.0 mag is clearly seen, although it exhibits a gap of $\Delta(C-T_1)$ $\sim$ 0.8 mag [$\Delta(B-V)$ = 0.41 mag] in the distribution of MS stars. Similar gaps occuring at nearly the same colour interval were previously noted in the CMDs of several known OCs, such as Praesepe (Johnson et al. 1962), NGC\,1039 (Canterna et al. 1979) and NGC\,2547 (Clari\'a 1982). The best isochrone comparison to the cluster CMD confirms that Ruprecht\,144, affected by a relatively large reddening [$E(B-V)$ = 0.70 mag], is a young cluster [log (age) = 8.0-8.2], located at 1.51 kpc from the Sun (Table 6). Both reddening and distance here derived show a reasonable agreement with the results obtained from 2MASS data (K13, Camargo et al. 2009), although we find a significantly younger age (Table 7) consistent with that derived from integrated spectroscopy (Ahumada et al. (2000). As in the case of ESO\,303-SC03, the large age difference between ours and that derived from 2MASS data may be due to a difference in the turn-off point adopted in the CMD. 

\subsection{Archinal\,1}
This likely cluster, also listed as MWSC\,3021 (K12), was first noted by AH03 who described it as a Trumpler (1930) type II2p cluster in Serpens containing about 24 stars within an area of 1.5' in diameter. There are 4 GSC stars included in the cluster field, the brightest of which is GSC\,0045700629 (Fig. 3). The presence of the cluster is clearly seen in the RDP {\bf (Fig. 5)}, from which we estimate a small angular radius of 1.4', almost twice the value estimated by AH03. The cluster and the field CMDs are different {\bf (Fig. 9)}. However, as in Ruprecht\,87, a clear sequence is not observed in the cleaned CMD thus suggesting that we are not dealing with a physical system.

\subsection{Berkeley\,82}
This small-sized cluster (C1909+129), also named MWSC\,3065 by K12, was first identified by Setteducati and Weaver (1962) in the Aquila constellation. As indicated by its Trumpler class (III-1p), this poorly-populated cluster does not show a strong central concentration but it clearly stands out by its relatively dense population compared to that of the field stars (Fig. 3). According to Lyng\aa\,(1987) and AH03, it has an angular radius of 1.2', somewhat lower than the $r_{cl}$ value here derived (Table 5). As most of the clusters here studied, the presence of Berkeley\,82 is clearly seen in the RDP {\bf (Fig. 5)}. The cluster cleaned CMD has a well-defined MS, when compared to the field CMD {\bf (Fig. 9)}. The cluster parameters are estimated based on the MS and its apparent turn-off. The cleaned CMD of the cluster is well matched by the isochrones corresponding to a young cluster [log (age) = 7.9-8.0], affected by $E(B-V)$ = 0.85 mag, and located at a heliocentric distance of 1.82 kpc. From photoelectric UBV photometry, Forbes (1986) estimated nearly the same age within errors, but a somewhat larger reddening of $E(B-V)$ = 1.01 mag and a smaller distance of  0.98 kpc. 

\section{Conclusions}

We present homogeneous CT$_1$ CCD photometry in the field of 10 low Galactic latitudes OCs (or candidates) with the goal of estimating their fundamental parameters. In five cases (ESO\, 313-SC03, ESO\,129-SC32, Basel\,5, Ruprecht\,144 and Archinal\,1), we report results derived from the first CCD photometric study in the optical spectral range. We performed a star count analysis of the observed fields to assess the clusters' reality as over-densities of stars with respect to the field stellar background, and to estimate their apparent angular and linear radii. After applying a statistical subtraction method, we built field star decontaminated CMDs for our cluster sample. Ruprecht\,87 and Archinal\,1 are found to be probably not physical systems. The derived parameters for the remaining 8 studied OCs are presented in Table 6, together with their corresponding uncertainties. In all cases, the best fits were obtained using nearly solar metallicity isochrones. We also list in Table 6 the Galactocentric rectangular coordinates X$_\odot$, Y$_\odot$ and Z$_\odot$ and the Galactocentric cluster distances $R_{GC}$, derived by assuming 8.5 kpc for the Sun's Galactocentric distance. Eight of the cluster candidates studied here are located in the first and fourth quadrants, all of them inside the solar circle. Collinder\,347 was found to be the youngest object of our sample with an age of around 5 Myr, while Basel\,5 ($\sim$ 1 Gyr) is the oldest one. Five clusters (ESO\,313-SC03, ESO\,129-SC32, BH\,217, Collinder\,347 and Basel\,5) are found to be affected by differential reddening, while BH\,217 appears to be the most heavily reddened with a mean colour excess $E(B-V)$ = 1.30 mag. In general terms, the results obtained show fairly good agreement with previous photometric results. In ESO\,303-SC03, ESO\,129-SC32 and Basel\,5, however, our results are not consistent with those derived using 2MASS photometry. According to Dias et al. (2012), accuracy in the determination of the colour excess $E(J-H)$ using only 2MASS data is limited mostly by structural inaccuracy in the MS and/or in the MS narrow range of magnitude sampling. It is probable that the main source of this disagreement is the limiting magnitude shown in the sampling of the MS, together with photometric errors. Note that 2MASS photometric errors typically reach 0.10 mag at J $\leq$ 16.2 mag and H $\leq$ 15.0 mag (Soares and Bica 2002), while in optical bands (CT$_1$, for example), they are usually lower than 0.05 mag at V $\leq$ 18.0 mag and T$_1$ $\leq$ 19.5 mag. The present study provides new OC parameters and some revisions to the Dias et al. (2002) catalogue.

\MakeTable{|l|c|c|c|c|c|c|}{12.5cm}{Basic parameters of the selected clusters}
{\hline
Cluster &   $\alpha$$_{\rm 2000}$  &  $\delta$$_{\rm 2000}$  &   {\it l} &    b      &   D  &  Trumpler class \\
        &     (h:m:s)              & ($^{\circ}$ ' ")         & ($^{\circ}$)  &  ($^{\circ}$) & (')  &                  \\
\hline
ESO\,313-SC03 &  08:31:42   &       -41:47:00   &     260.43    &     -1.31    &      2.5    &         II3    \\               
BH\,54	      &  08:49:41   &       -44:21:12   &     264.48    &     -0.28    &      4.0    &         II1p   \\                
Ruprecht\,87  &  10:15:30   &       -50 42:18   &     279.37    &      4.88    &      4.0    &        III2p   \\      
ESO\,129-SC32 &	 11:44:12   &       -61:05:00   &     294.89    &      0.76    &      4.0    &         II2    \\
BH\, 217      &  17:16:20   &       -40:49:50   &     346.78    &     -1.51    &      5.0    &          I2m   \\      
Collinder\,347&	 17:46:18   &       -29:20:00   &     359.74    &     -0.33    &     10.0    &         II2m,n \\
Basel\,5      &  17:52:28   &       -30:06:23   &     359.77    &     -1.87    &      5.0    &        III2mn  \\	   
Ruprecht\,144 &  18:33:33   &       -11:25:11   &      20.75    &     -1.28    &      2.0    &         IV1p   \\
Archinal\,1	&18:54:49   &	    +05:32:56   &     38.26     &      1.78    &      1.5    &         II2p   \\
Berkeley\,82	&19:11:24   &	    +13:04:00   &     46.82     &      1.59    &      2.5    & 	      III1p   \\
\hline
}

\MakeTable{|l|c|c|c|c|}{12.5cm}{Observation log of observed clusters}
{\hline
Cluster       &	     Date     &	Filter &   Exposure time & Airmass    \\
              &               &        &  (sec)          &            \\
\hline
ESO\,313-SC03   &   May 11, 2008       &     C  &          60, 90, 900               &   1.17    \\
                &                      &     R  &       25, 25, 150, 200             &   1.22    \\
BH\,54		&   May 11, 2008       &     C  &           90, 90, 450              &   1.23    \\
                &                      &     R  &          5, 10, 15                 &   1.23    \\ 
Ruprecht\,87	&   May 11, 2008       &     C  &         60, 60, 600, 600           &   1.25    \\ 
                &                      &     R  &         5, 10, 10, 60              &   1.25    \\        
ESO\,129-SC32   &   May 11, 2008       &     C  &        60, 60, 450, 450            &   1.28    \\ 
                &                      &     R  &         10, 10, 60, 120            &   1.26    \\
BH\,217         &   May 11, 2008       &     C  &         45, 70, 600, 600           &   1.08    \\ 
                &                      &     R  &      10, 20, 30, 150, 150          &   1.08    \\
Collinder\,347  &   May  11, 2008      &     C  &         30, 30, 400, 400           &   1.01    \\ 
  		&	               &     R  &         15, 15, 150, 150           &   1.01    \\
Basel\,5	&   May 11, 2008       &     C  &          30, 30, 600, 600          &   1.00    \\
                &                      &     R  &            15, 15, 120, 120        &   1.00    \\
Ruprecht\,144	&   May 11, 2008       &     C  &         30, 30, 300, 300           &   1.06    \\
		&		       &     R  &          20, 20, 150, 150          &   1.05    \\
Archinal\,1	&   May 9, 2008	       &     R  &             7, 45   	             &   1.23    \\
  		&		       &     C  &	         45, 450             &   1.24    \\
Berkeley\,82	&   May 11, 2008       &     C  &                 30, 60, 450        &   1.38    \\ 
		&		       &     R  &        5, 5, 15, 20, 120, 120, 450 &   1.39    \\ 
\hline                                                                                        
}       

\MakeTable{|c|c|}{12.5cm}{Standard system mean calibration coefficients}
{\hline
C     &   $T_1$   \\
\hline
$a_1$ = 3.61 $\pm$ 0.03   &  $b_1$ = 3.04 $\pm$ 0.02  \\
$a_2$ = 0.56 $\pm$ 0.01   &  $b_2$ = 0.33 $\pm$ 0.02  \\
$a_3$ = -0.19 $\pm$ 0.01 & $b_3$ = -0.03 $\pm$ 0.01 \\
\hline
}

\MakeTable{|c|c|c|c|c|c|c|c|c|c|}{12.5cm}{CCD $CT_1$ data of stars in the field of ESO\,313-SC03}
{\hline
Star & $X$\hspace{0.2cm} &  $Y$\hspace{0.2cm} &   $\alpha$$_{\rm 2000}$  &  $\delta$$_{\rm 2000}$  & $T_{1}$  & $\sigma$$T_1$  & $C-T_1$  & $\sigma$$(C-T_1)$ & n \\
     & (pixel)           & (pixel)            &     ($^{\circ}$)              & ($^{\circ}$)            &  (mag)   & (mag)          &  (mag)   & (mag)           &  \\
\hline
1  &   134.274   &  4.971  & 128.023 & -41.664 &  17.169 &  0.036  & 1.940  & 0.045  & 1  \\
2  &   329.086   &  6.259  & 127.994 & -41.665 &  15.370 &  0.040  & 1.909  & 0.057  & 2  \\
3  &   772.637   &  7.016  & 127.923 & -41.665 &  18.524 &  0.023  & 2.357  & 0.098  & 1  \\
\hline
}

\MakeTable{|l|c|c|c|c|}{12.5cm}{Cluster sizes}
{\hline
Cluster         & \multicolumn{2}{c}{r$_{FWHM}$} &    \multicolumn{2}{c}{r$_{cl}$} \\
                & (')         & (pc)             &      (')      & (pc) \\
\hline
ESO\,313-SC03   &       0.4   &    0.6           &        2.1   &  3.2     \\ 
BH\,54          &       0.5   &    0.1           &        2.1   &     0.6     \\
Ruprecht\,87    &       0.4   &     -            &         1.7   &       -     \\
ESO\,129-SC32   &        -    &    -             &      2.0$^a$  &      1.2    \\
BH\,217         &       0.5   &   0.3            &         3.4   &      2.0    \\
Collinder\,347  &       2.5   &   1.3            &         5.5   &  2.8    \\
Basel\,5        &       0.3   &   0.2            &         2.4   &      1.8    \\
Ruprecht\,144   &       0.7    &   0.3            &         1.0   &       0.4   \\
Archinal\,1     &       0.4   &   -              &         1.4   &      -      \\
Berkeley\,82    &       1.0   &   0.5            &         1.8   &      1.0    \\
\hline 
\multicolumn{5}{p{4cm}}{$^a$Taken from AH03}
}

\MakeTable{|l|c|c|c|c|c|c|c|}{12.5cm}{Fundamental parameters estimated for the clusters}
{\hline
Cluster  &  $E(B-V)$  &  $d$     &     Age  &     $X$    &  $Y$    &  $Z$    & $R_{GC}$   \\
         &   (mag)    &  (kpc)   &    (Myr) &     (kpc)  &  (kpc)  &  (pc)  &  (kpc)      \\
\hline
ESO\,313-SC03  &    0.80$\pm$0.10  &      5.01$\pm$0.57    &           400$(+100/-80)$   &        -4.9   &      0.8  &     -100   &     10.6        \\ 
BH\,54         &    0.75$\pm$0.03  &      1.00$\pm$0.03    &            30$(+70/-20)$    &        -1.0   &      0.1  &       -10  &      8.7        \\
ESO\,129-SC32  &    0.10$\pm$0.09  &      2.00$\pm$0.22    &           400$(+100/-80)$   &        -1.8   &    -0.8   &        30  &      7.9        \\
BH\,217	       &    1.30$\pm$0.19  &      2.09$\pm$0.45    &              40$(+10/-10)$   &        -0.5   &    -2.0   &       -60  &      6.5        \\
Collinder\,347 &    0.80$\pm$0.17  &      1.74$\pm$0.34    &                5$(+5/-4)$   &         0.0   &    -1.7   &       -10  &      6.8        \\
Basel\,5       &    0.48$\pm$0.15  &      2.51$\pm$0.43    &          1000$(+300/-200)$  &         0.0   &    -2.5   &       -80  &      6.0        \\
Ruprecht\,144  &    0.70$\pm$0.03  &      1.51$\pm$0.05    &            130$(+30/-20)$   &         0.5   &    -1.4   &       -30  &      7.1        \\
Berkeley\,82   &    0.85$\pm$0.05  &      1.82$\pm$0.04    &              80$(+10/-10)$   &         1.3   &   -1.2    &      50    &      7.4        \\
\hline    
}

\MakeTable{|l|c|c|c|c|}{12.5cm}{Previous estimates of cluster parameters}
{\hline
Cluster  &  $E(B-V)$  &  $d$     &     Age  &     References$^b$ \\
         &   (mag)    &  (kpc)   &    (Myr) &                \\
\hline
ESO\,313-SC03  & 0.19  &        3.71   &     2800   &    (1)   \\
               & 1.08  &        4.27   &      398   &    (2)   \\
BH\,54         & 0.60  &        1.10   &        7   &    (1)   \\
 	       & 0.73  &        1.24   &       32   &    (2)   \\
	       & 0.75  &        1.30   &       60   &    (3)   \\
ESO\,129-SC32  & 0.42  &        1.56   &      398   &    (2)   \\
BH\,217	       & 1.15  &        1.69   &       45   &    (1)   \\
               & 1.30  &        1.72   &       50   &    (2)   \\
               & 0.80  &         -     &       35   &    (4)   \\
               & 1.00  &        1.68   &       56   &    (5)   \\
               & 0.80  &        1.21   &       35   &    (8)   \\
Collinder\,347 & 1.02  &        1.41   &       13   &    (1)   \\
               & 1.44  &        1.85   &        4   &    (2)   \\
               & 1.16  &        1.51   &       -    &    (6)   \\
Basel\,5       & 1.85  &        1.08   &     1580   &    (1)   \\
               & 0.34  &        1.00   &      708   &    (2)   \\
Ruprecht\,144  & 0.67  &        1.65   &      624   &    (2)   \\
               & 0.32  &         -     &      150   &    (4)   \\
               & 0.77  &        1.6    &      450   &    (9)  \\
Berkeley\,82   & 0.92  &        1.13   &      115   &    (2)   \\
               & 1.01  &        0.98   &       70   &    (7)   \\
\hline
\multicolumn{5}{p{9cm}}{$^b$ (1) B11; (2) K13; (3) Piatti et al. (2010); (4) Ahumada et al. (2000); (5) Caetano et al. (2015); (6) Moffat and Vogt (1975); (7) Forbes (1986); (8) McSwain and Gies (2005); (9) Camargo et al. (2009)}
}

\begin{figure}
\includegraphics[width=\textwidth]{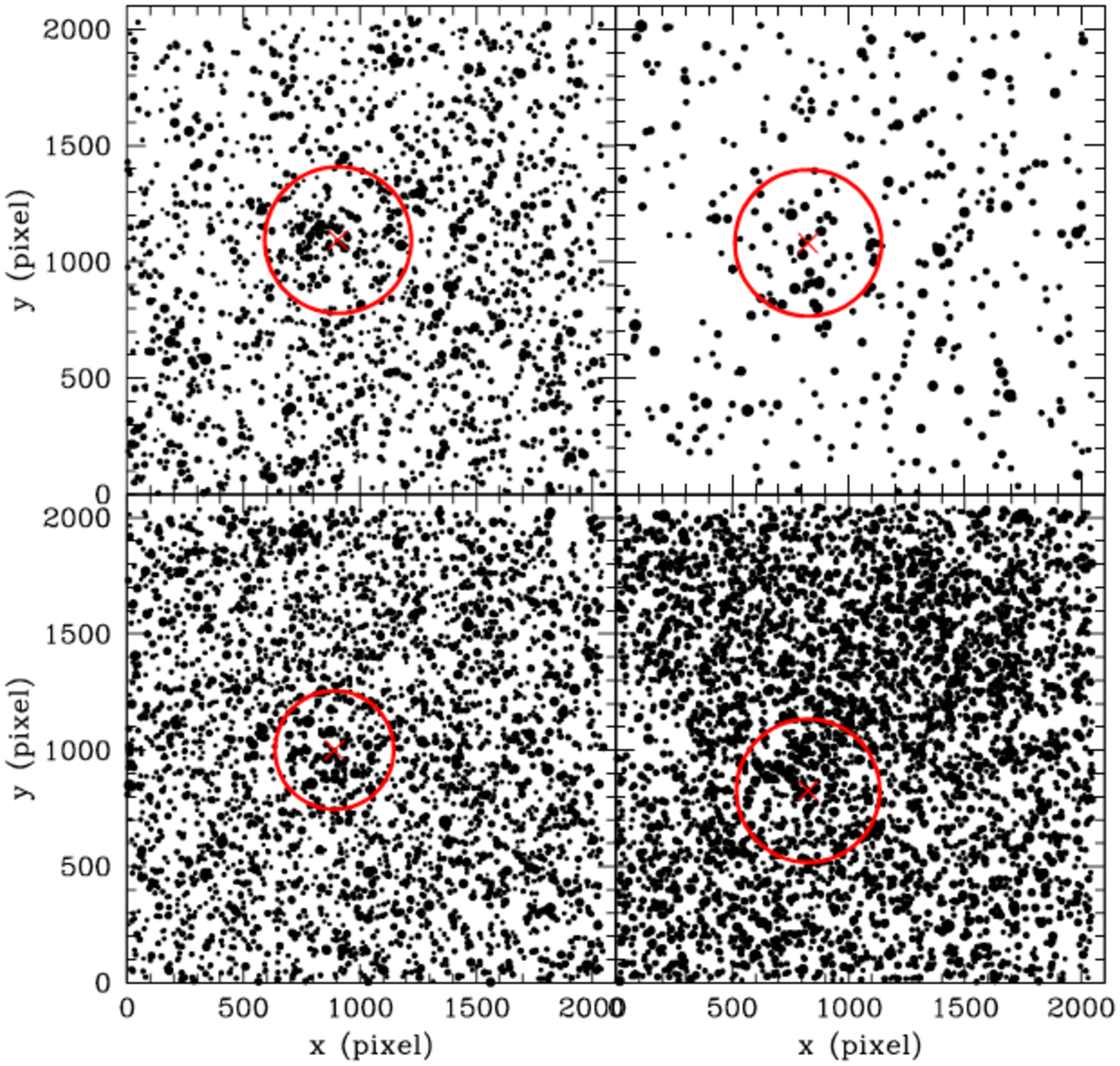}
\FigCap{Schematic finding charts of the stars observed in ESO\,313-SC03 (top left), BH\,54 (top right), Ruprecht\,87 (bottom left) and ESO\,129-SC32 (bottom right). North is up and East is to the left. The sizes of the plotting symbols are proportional to the T$_1$ brigthness of the stars. Red circles r$_{cl}$ wide are shown around the adopted cluster centres (crosses).}
\end{figure}

\begin{figure*}
\includegraphics[width=\textwidth]{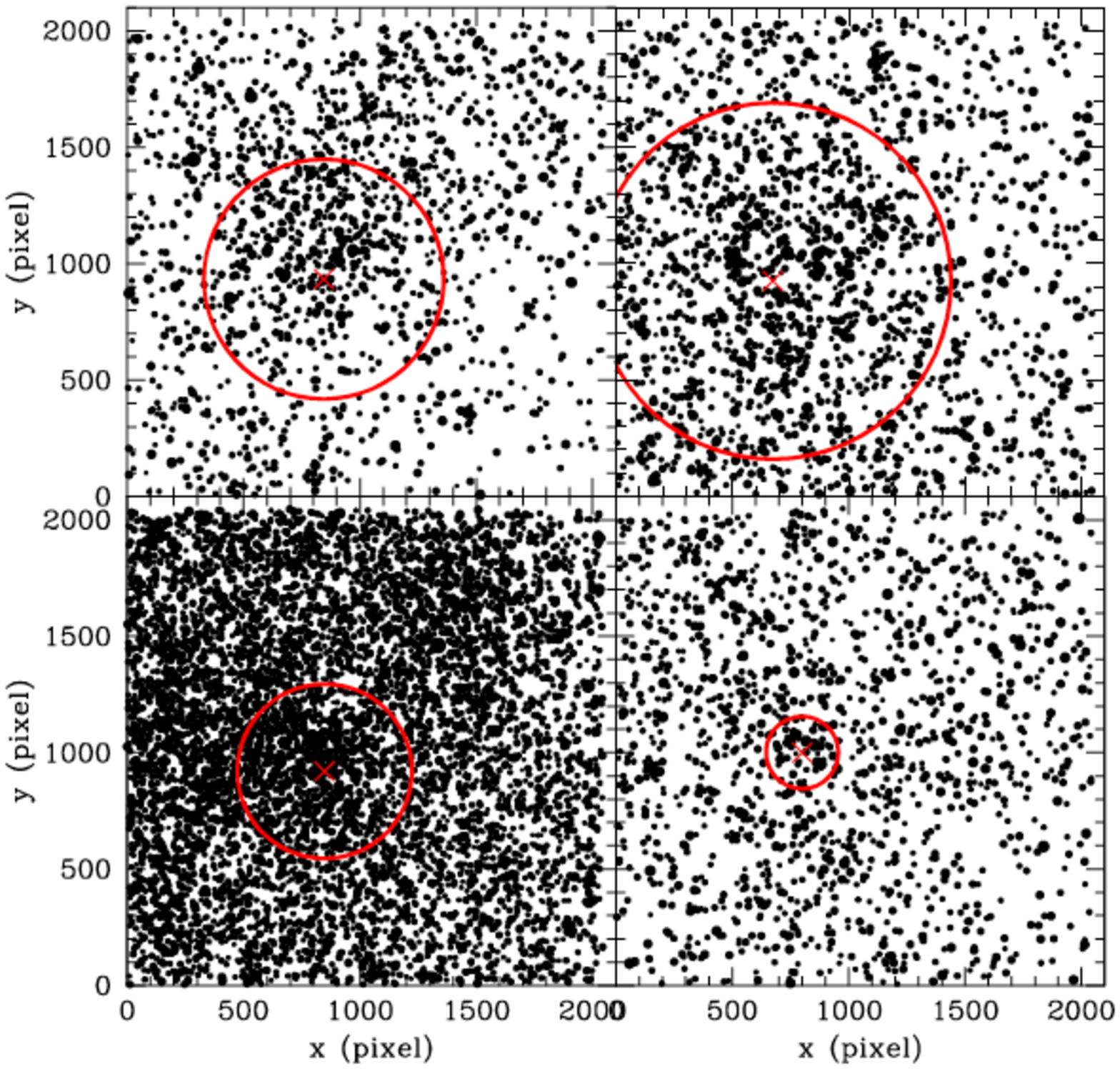}
\caption{Same as Fig. 1 but for the stars observed in BH\,217 (top left), Collinder\,347 (top right), Basel\,5 (bottom left) and Ruprecht\,144 (bottom right)}
\label{fig:2}
\end{figure*}

\begin{figure*}
\includegraphics[width=\textwidth]{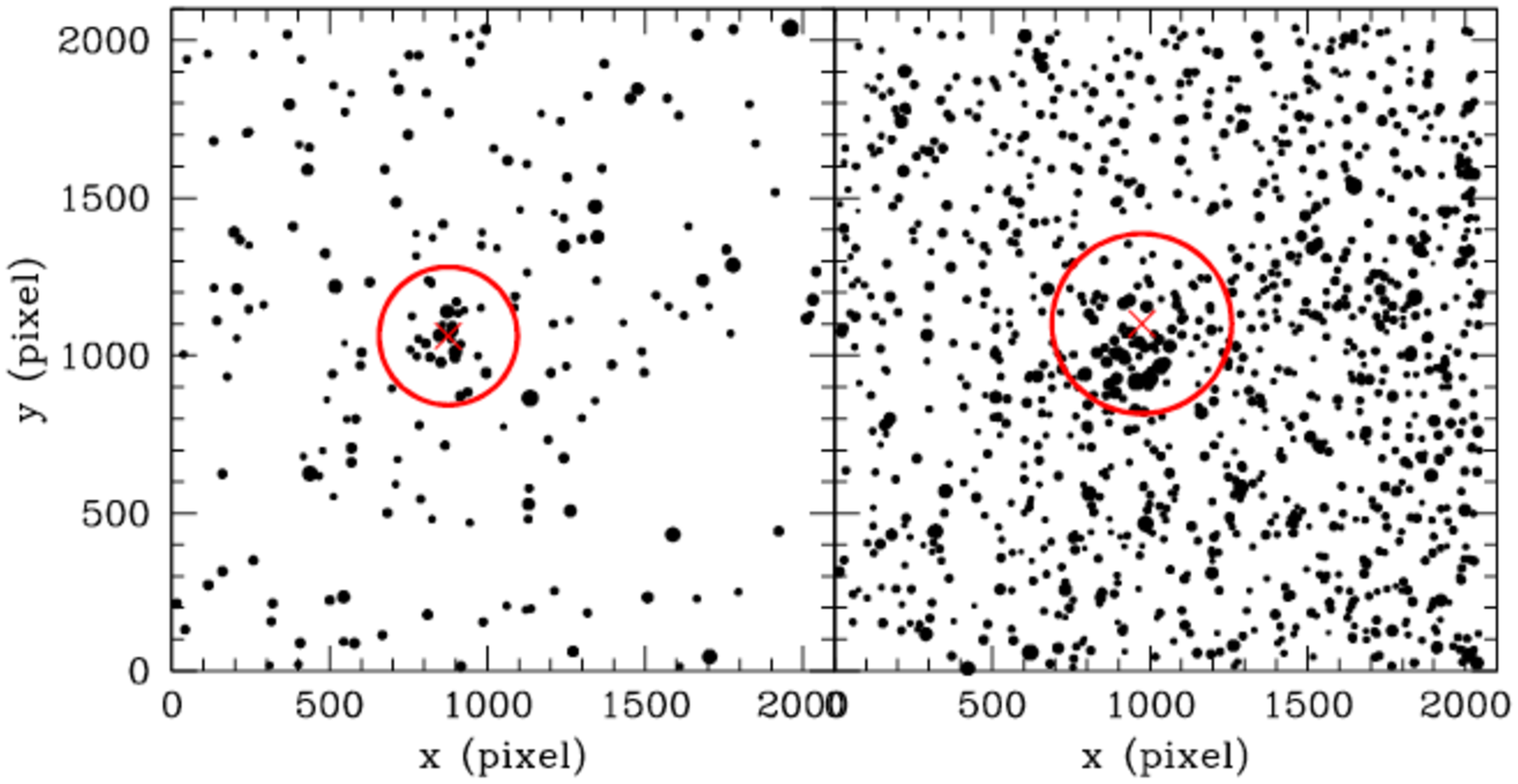}
\caption{Same as Fig. 1 but for the stars observed in Archinal\,1 (left panel) and Berkeley\,82 (right panel)}
\label{fig:3}
\end{figure*}

\begin{figure*}
\includegraphics[width=\textwidth]{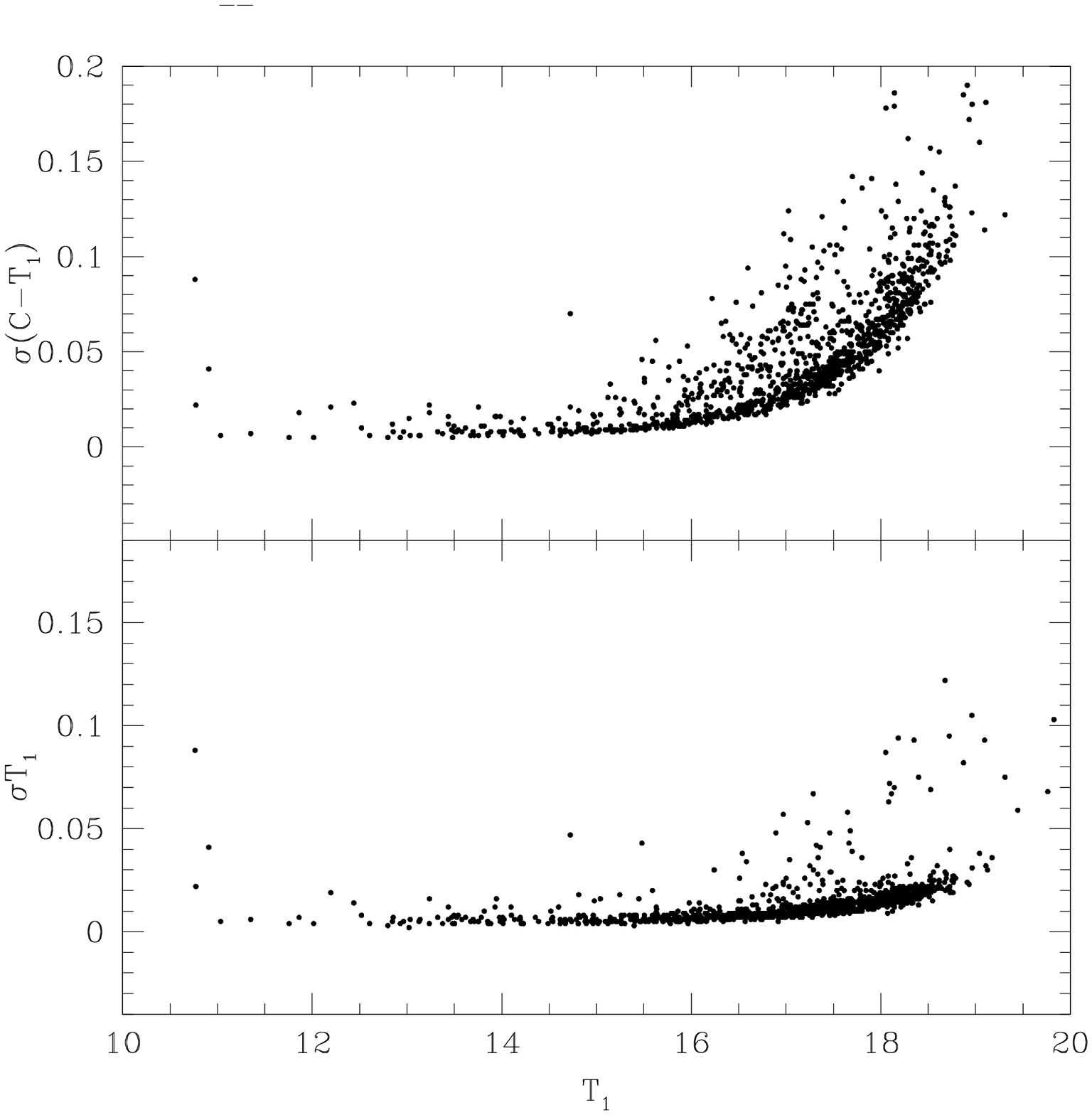}
\caption{T$_1$ magnitude and C-T$_1$ colour photometric errors as a function of T$_1$ for stas measured in the field of Berkeley\,82}
\label{fig:4}
\end{figure*}

\begin{figure*}
{\includegraphics[width=\textwidth]{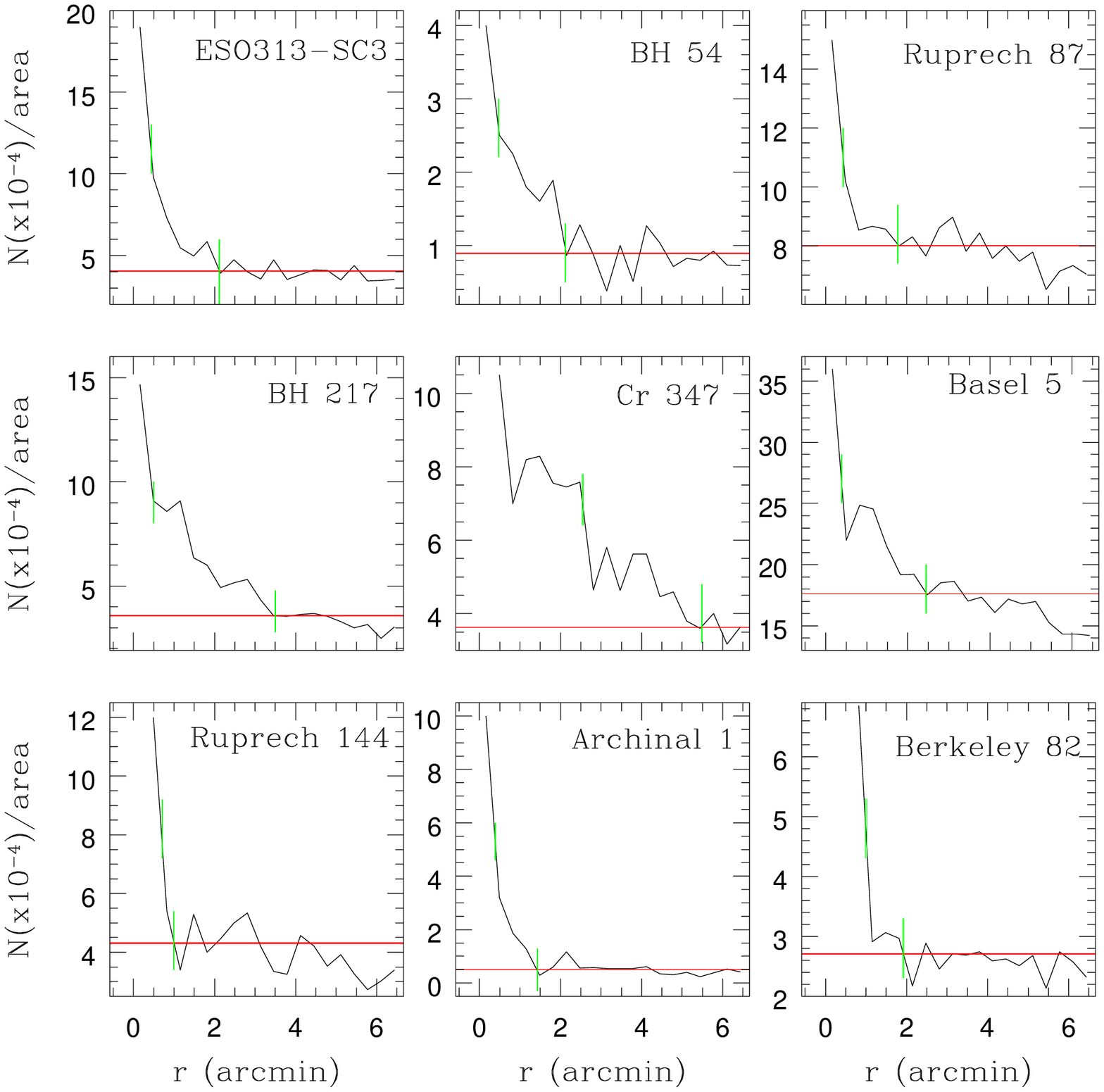}}
\caption{Radial number density profiles for the stars in the studied open clusters or candidates. As explained in Section 3.1, it was not possible 
to construct the RDP of ESO 129-SC32. The cluster names are indicated inside each panel. The red horizontal lines represent the background levels. 
The radius at FWHM (r$_{FWHM}$) and the adopted cluster radius (r$_{cl}$) are indicated by green vertical lines.} 
\label{fig:5}
\end{figure*}

\begin{figure*}
\label{fig:8}
\includegraphics[width=\textwidth]{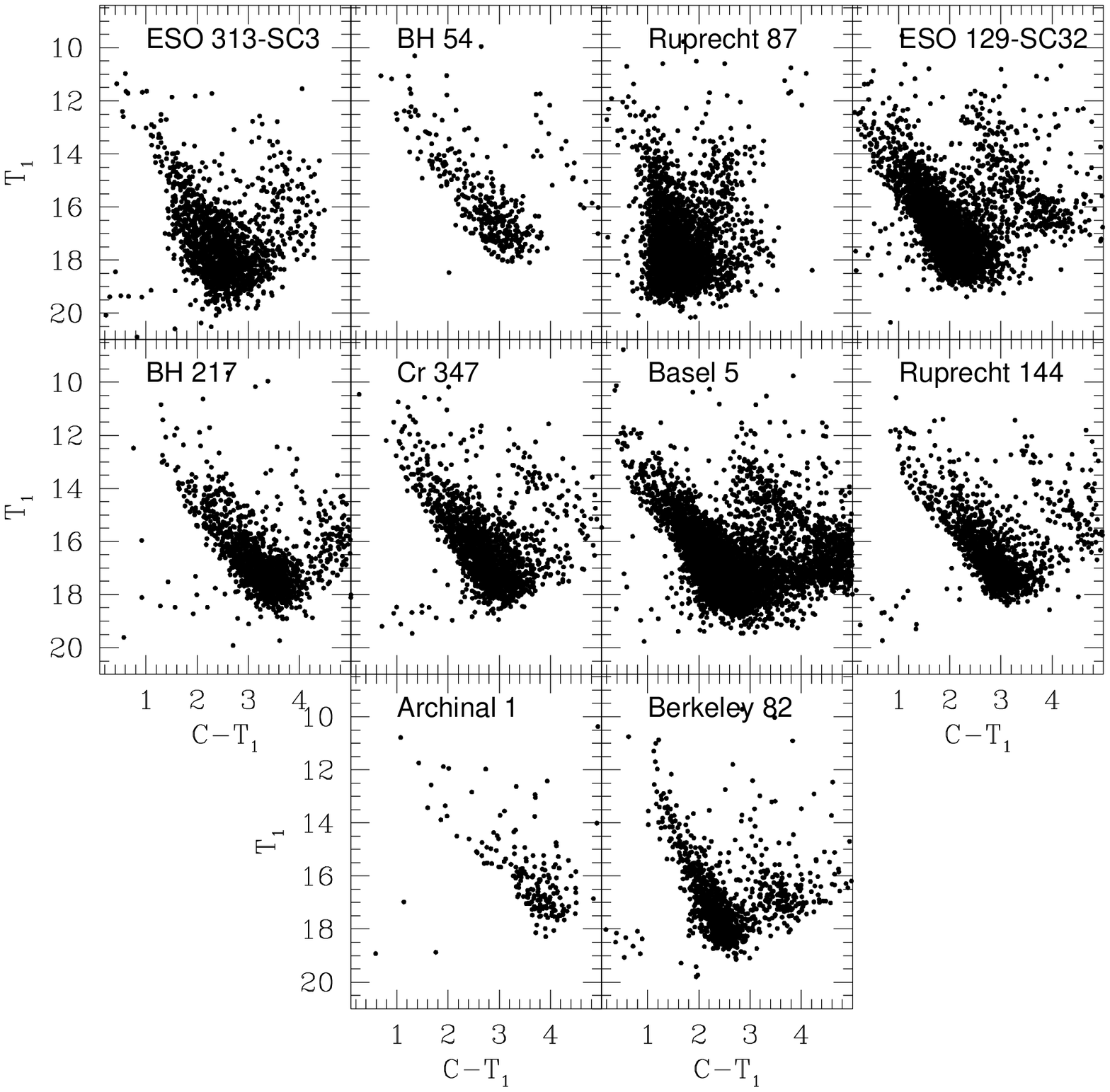}
\caption{(C-T$_1$,T$_1$) CMDs for stars observed in the 13.6'$\times$ 13.6' fields of the studied cluster candidates}
\end{figure*}

\begin{figure*}
\includegraphics[width=\textwidth]{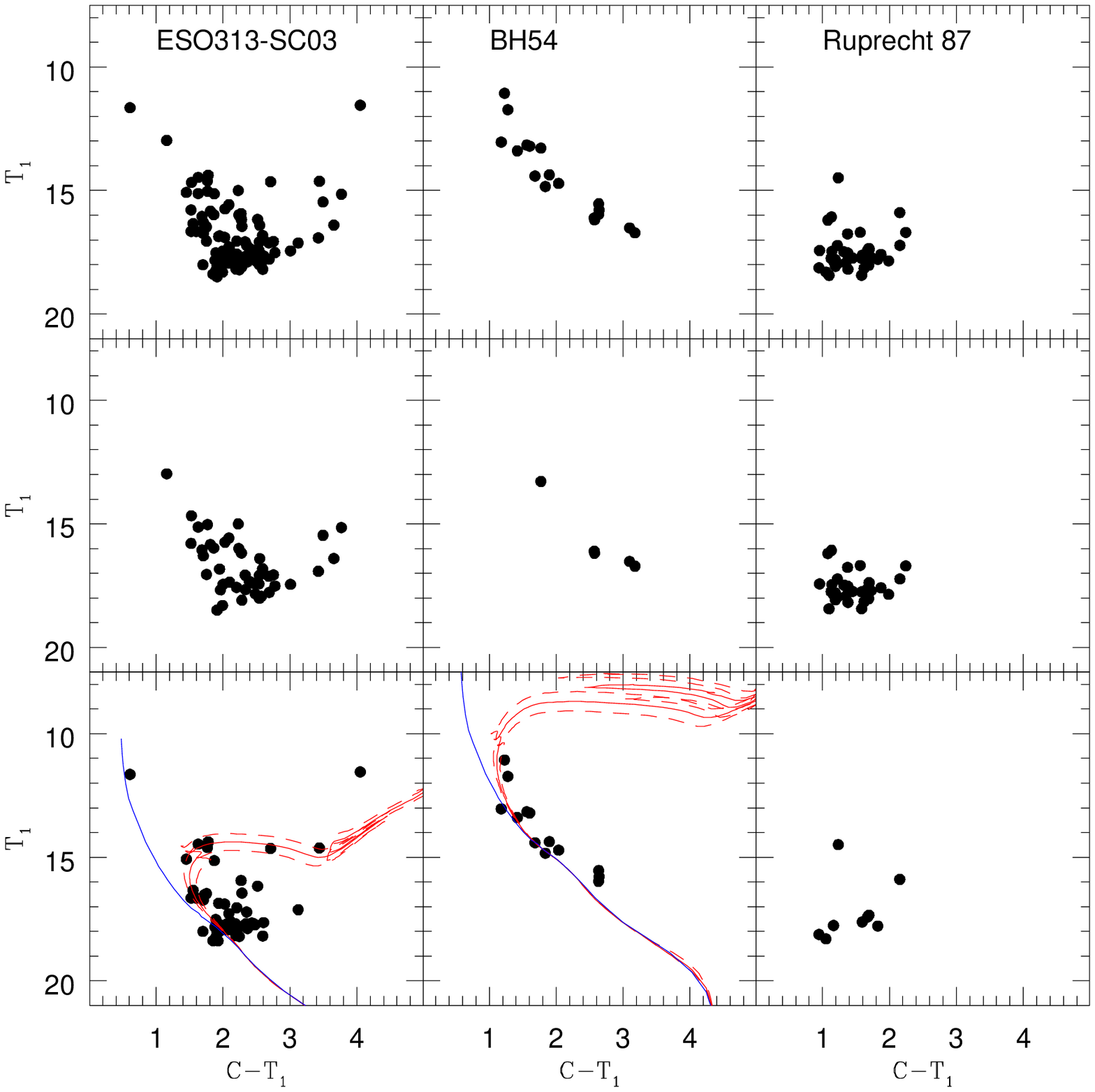}
\label{fig:9}
\caption{(C-T$_1$,T$_1$) CMDs for the clusters ESO\,313-SC03, BH\,54 and Ruprecht\,87. Top panels: observed CMDs including stars located within the adopted cluster radii (Table 5). Middle panels: equal area comparison fields. Bottom panels: field-star decontaminated CMDs wherein the ZAMS and adopted isochrones from Bressan et al. (2012) are overplotted with solid lines. The isochrones associated to the cluster age errors are indicated by dashed lines, for comparison purposes. Only stars with colour uncertainties smaller than 0.06 mag are included in the diagrams.}
\end{figure*}

\begin{figure*}
\label{fig:10}
\includegraphics[width=\textwidth]{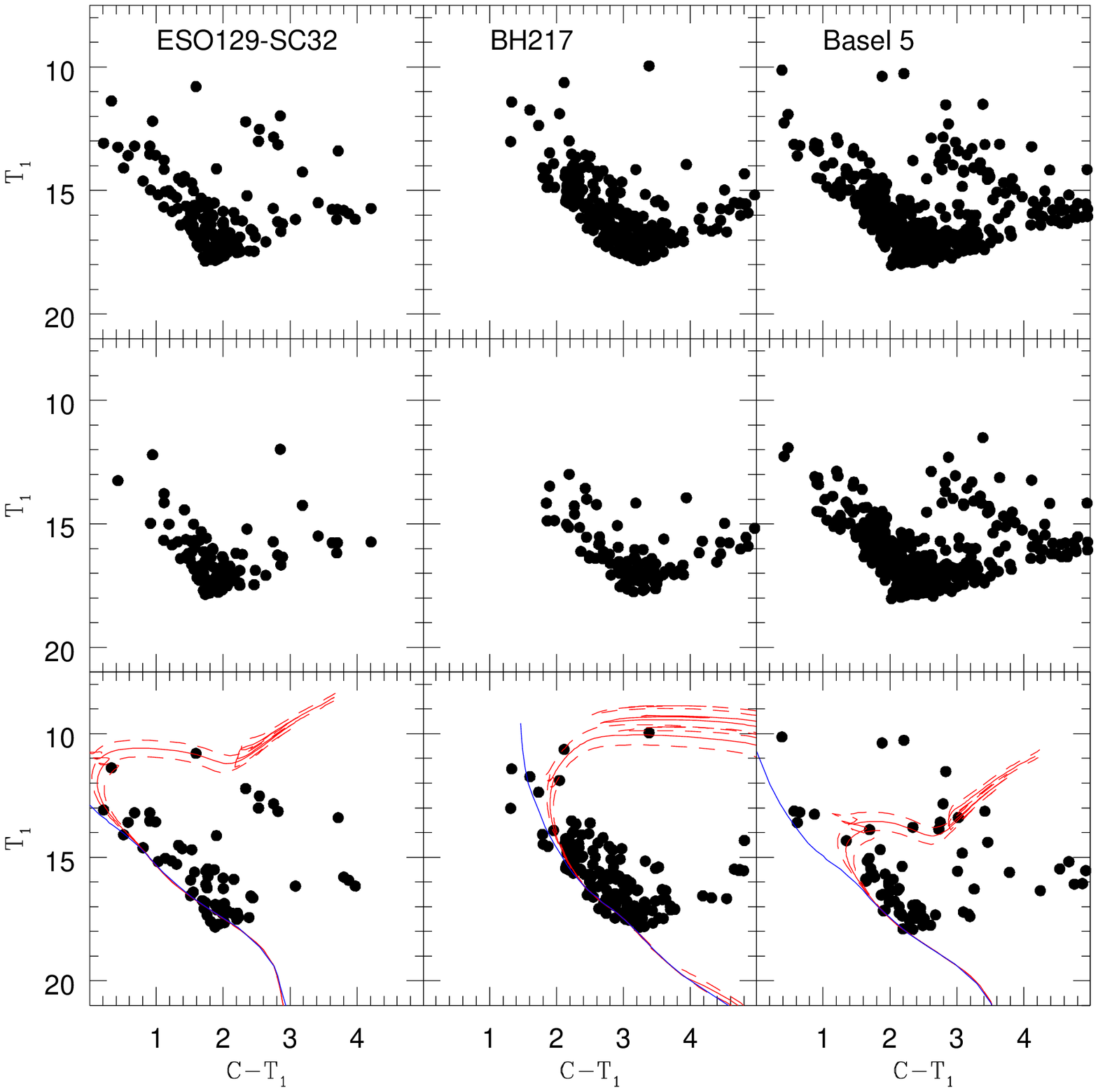}
\caption{Same as {\bf Fig. 7} but for the clusters ESO\,129-SC32, BH\,217 and Basel\,5}
\end{figure*}

\begin{figure*}
\includegraphics[width=\textwidth]{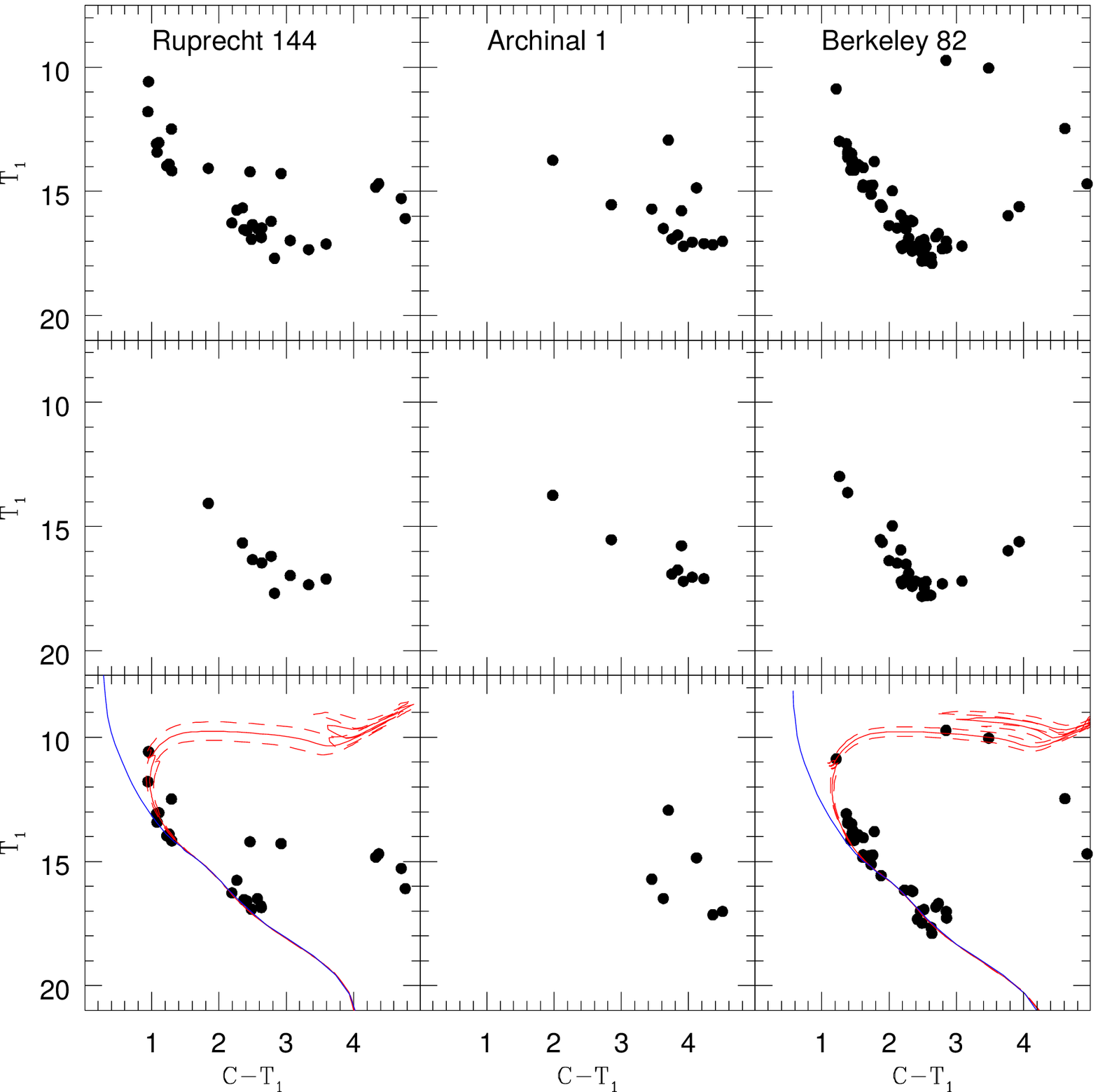}
\label{fig:11}
\caption{Same as {\bf Fig. 7} but for the clusters Ruprecht\,144, Archinal\,1 and Berkeley\,82}
\end{figure*}

\begin{figure*}
\includegraphics[width=\textwidth]{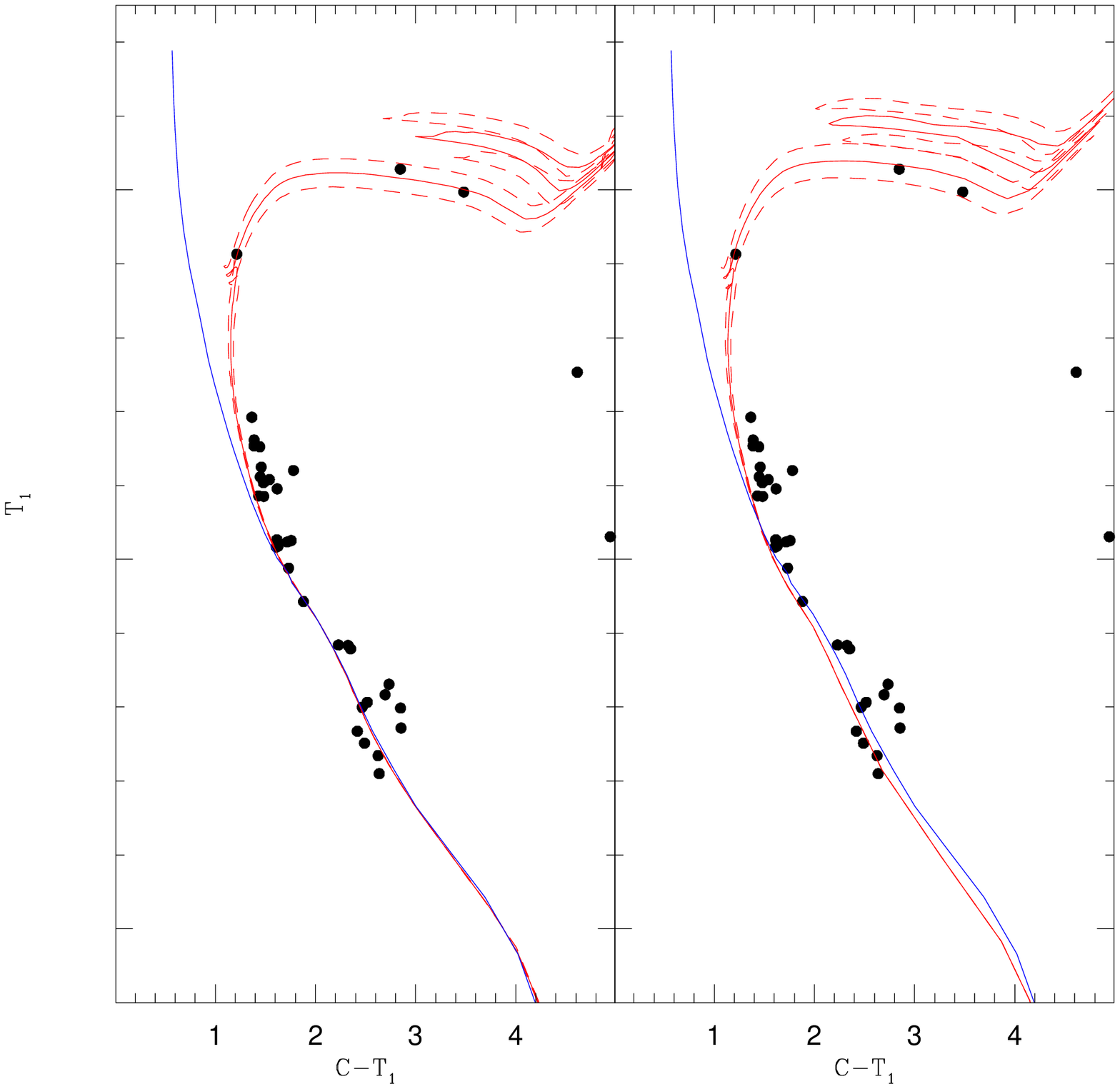}
\label{fig:12}
\caption{Field-star decontaminated CMDs of Berkeley 82 wherein the ZAMS and isochrones from Bressan et al. (2012) for Z = 0.019 (left panel) 
and Z = 0.013 (right panel) are overplotted with blue and red solid lines, respectively. Reddening, distance and age values used in the fittings 
are those listed in Table 6 for Berkeley 82. The isochrones associated to the cluster age errors are indicated by dashed red lines.}
\end{figure*}

\begin{figure*}
\includegraphics[width=\textwidth]{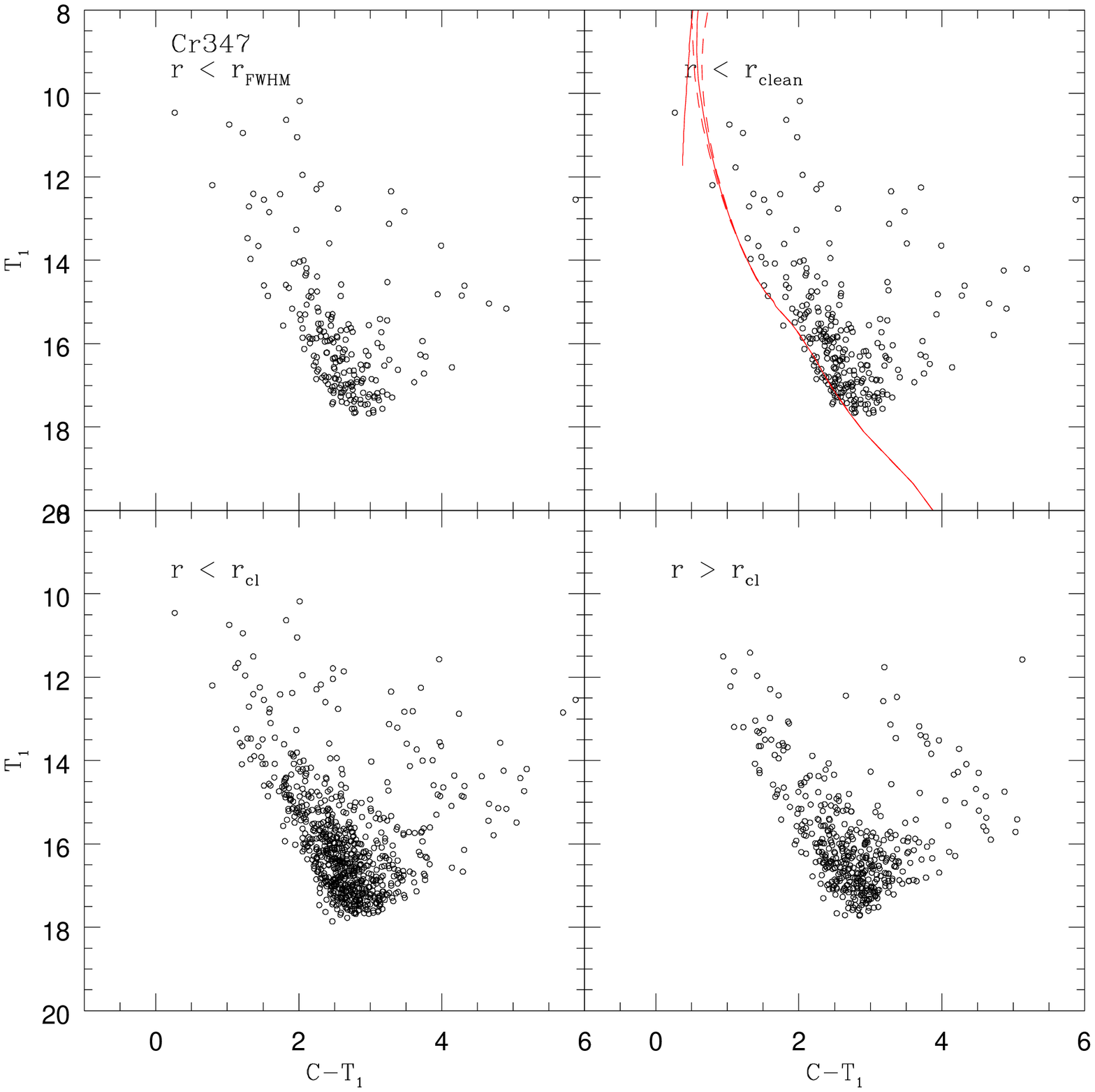}
\label{fig:12}
\caption{CMDs for stars observed in different extracted circular regions around Collinder\,347 centre as indicated in each panel}
\end{figure*}

\Acknow{J.J. Clari\'a, T. Palma and A.V. Ahumada are gratefully indebted to the CTIO staff for their kind hospitality and support during the observing run. This research was partially supported by the Argentinian institutions CONICET, SECYT (Universidad Nacional de C\'ordoba) and Agencia Nacional de Promoci\'on Cient\'{\i}fica y Tecnol\'ogica (ANPCyT). We have used both the SIMBAD database, operated at CDS, Strasbourg, France, and the NASA's Astrophysics Data System. This work is based on observations made at Cerro Tololo Inter-American Observatory, which is operated by AURA, Inc., under cooperative agreement with the NSF. We also made use of the VizieR catalogue access tool, CDS, Strasbourg, France. The original description of the VizieR service was published in A\&AS 143, 23.}

\end{document}